\pdfoutput=1
\documentclass[reprint,aps,pre,showpacs,twocolumn]{revtex4-1}
\usepackage{mathptmx}
\usepackage{epsfig,amsopn}
\usepackage{graphicx}
\usepackage{amsmath,amssymb}
\usepackage{natbib}
\usepackage{braket}
\usepackage{float}
\usepackage{enumerate}
\allowdisplaybreaks[1]

\newcommand{\eref}[1]{Eq.~(\ref{#1})}%
\newcommand{\Eref}[1]{Equation~(\ref{#1})}%
\newcommand{\fref}[1]{Fig.~\ref{#1}} %
\newcommand{\Fref}[1]{Figure~\ref{#1}}%
\newcommand{\sref}[1]{Sec.~\ref{#1}}%
\newcommand{\aref}[1]{Appendix~\ref{#1}}%

\def\bea{\begin{eqnarray}}
\def\eea{\end{eqnarray}}

\def\nn{\nonumber}

\def\f{\frac}

\begin{document}

\title{Work fluctuations for a Brownian
particle driven by a correlated external random force}

\author{Arnab Pal and Sanjib Sabhapandit}

\affiliation{Raman Research Institute, Bangalore 560080, India}

\date{\today}



\begin{abstract}
We have considered the underdamped motion of a Brownian particle in
the presence of a correlated external random force. The force is
modeled by an Ornstein-Uhlenbeck process. We investigate the
fluctuations of the work done by the external force on the Brownian
particle in a given time interval in the steady state. We calculate
the large deviation functions as well as the complete asymptotic form
of the probability density function of the performed work. We also
discuss the symmetry properties of the large deviation functions for
this system. Finally we perform numerical simulations and they are in
a very good agreement with the analytic results.
\end{abstract}

\maketitle

\section{Introduction}
\label{sec:Introduction}

In recent times the Fluctuation Theorem (FT) has generated a lots of
excitement in the field of non-equilibrium statistical mechanics, as
it allows thermodynamic concepts to be applied to also small systems,
as well as to systems that are arbitrarily far from equilibrium.
The FT expresses universal properties of the probability density
function (PDF) $p(\Omega)$ for functional $\Omega[x(\tau)]$, like
work, heat, power flux or entropy production, evaluated along the
fluctuating trajectories $x(\tau)$ taken from ensembles with
well-specified initial distributions.  There have been a number of
theoretical~\cite{Evans:94, Evans:93, Gallavotti:95, Gallavotti:96,
Kurchan:98, Lebowitz:99, Farago:02, vanZon:03, vanZon:04, Mazonka:99,
Narayan:04, Seifert:05, Baiesi:06, Bonetto:06, Visco:06,
Saito:07, Harris:07,Derrida:07, Kundu:11, Saito:11, Sabhapandit:11-12,
Arnab:13, Park:13, Gatien:13, Gatien:14} and
experimental~\cite{Wang:02, Wang:05, Carberry:04, Goldburg:01,
Liphardt:02, Collin:05, Majumdar:08, Douarche:06, Ciliberto:10-2,
Gomez-Solano:11, Ciliberto:2013} studies to elucidate different
aspects of FT.  We refer to the recent review \cite{Seifert:2012}
which contains an extensive list of references both from the
theoretical and the experimental aspects.

The FT can be broadly classified into two groups, namely, the
Transient FT (TFT) and the Steady State FT (SSFT).  The TFT pioneered
by Evans and Searles \cite{Evans:94} applies to relaxation towards a
steady state but at finite time. In this work, they obtain the
symmetries of the PDF of `Entropy Production' at the transient.  On
the other hand the SSFT quantifies the `Entropy Production'
$\Omega_{\tau}$ in a time duration $\tau$, in the non equilibrium
steady state as,
\begin{equation}
\f{p(\Omega_{\tau}=\omega \tau)}{p(\Omega_{\tau}=-\omega \tau)}\sim
e^{\tau \omega}~.
\label{Fluctuation Theorem} 
\end{equation}
This was first found by Evans \textit{et al.} in simulations of
two-dimensional sheared fluids~\cite{Evans:93} and then proven by
Gallavotti and Cohen~\cite{Gallavotti:95,Gallavotti:96} using
assumptions about chaotic dynamics. Kurchan \cite{Kurchan:98} and
Lebowitz and Spohn \cite{Lebowitz:99} have established this theorem
for stochastic diffusive dynamics.  In all these early works, the
entropy production has been identified with the entropy production in
the medium.  However, it was shown in \cite{Seifert:05} that the SSFT
holds even for finite times in the steady state if one incorporates
the entropy production of the system.  Though the FT for entropy
production has been found to be robust under rather general
conditions, the question is whether this is generic for other
observables like work, dissipated heat etc.  Indeed, there are only a
handful of examples where the SSFT for work, heat
\cite{Farago:02,vanZon:03, vanZon:04,
Mazonka:99, Baiesi:06, Bonetto:06, Visco:06, Saito:07, Kundu:11,
Saito:11, Sabhapandit:11-12, Arnab:13, Gatien:13, Gatien:14, Park:13} has been
investigated. It has been observed that the validation of SSFT for
these observables is not universal, e.g. in \cite{vanZon:03}, the
authors have found the `work' to satisfy SSFT while the `heat' does
not, in general.  Thus, one hopes to gain insights by studying exactly
solvable cases.

As the FT deals only with the symmetry properties of the PDF, the
explicit form of the PDF is often not required to realize the validity
of the relation \eqref{Fluctuation Theorem}.  However, it is by
itself, an interesting endeavor to compute the PDF of the time
integrated quantities like work, heat, etc., and there are not many
such examples where it can be done analytically.  The long time
behavior of the PDF is intimately related to the so-called large
deviation function (LDF)~\cite{Touchette:09}, and in the recent years,
a lot of efforts have been devoted to the computation of LDFs in
non-trivial models~\cite{Derrida:07, Harris:07}.  The symmetry
relation \eqref{Fluctuation Theorem} can be expressed in terms of a
symmetry relation satisfied by the corresponding LDF.

In this paper, we consider an underdamped Brownian particle driven by
a correlated random external field. We study the PDF of the work done
by the external random field in a given duration.  The exact LDF
associated with the PDF is found to have a non-trivial form.  The SSFT
is found to be hold in a restrictive parameter space of the model,
confirming the fact that the FT for work, heat is non generic.

The paper is organized as follows.  In the following section, we
define the model.  In \sref{sec:MGF} we compute the moment generating
function (MGF) of work $W_\tau$ performed in a given time $\tau$ in
steady state, which has the form $\langle e^{-\lambda
W_{\tau}}\rangle \sim g(\lambda) e^{\tau \mu(\lambda)}$.
In \sref{sec:PDF}, we invert the MGF to obtain the asymptotic form
(for large $\tau$) of the PDF of the work. We discuss the symmetry
properties of the large deviation functions and its connection with
the FT in \sref{sec:FT}. Finally we conclude
in \sref{sec:Summary}. Some details of the calculation has been
relegated to \aref{sec:Full-MGF}.


\section{Model}
\label{sec:Model}

Consider a Brownian particle of mass $m$, in the presence of an
external fluctuating time dependent field, at a temperature $T$.  The
velocity $v(t)$ of the particle evolves according to the underdamped
Langevin equation, given by,
\begin{equation}
m\frac{dv}{dt}+\gamma v=f(t) + \eta_{1}~, 
\label{Langevin-1}
\end{equation}
where $\gamma$ is the friction coefficient.  The viscous relaxation
time scale for the particle is $\tau_{\gamma}=m/\gamma$. The thermal
noise $\eta_{1}$ is taken to be a Gaussian white noise with mean zero
and correlation$\langle \eta_{1}(t)\eta_{1}(s)\rangle=2D\delta(t-s)$,
where diffusion constant $D=\gamma k_{B} T$ and $k_{B}$ is the
Boltzmann constant.  The external stochastic field $f$ is modeled by
an Ornstein-Uhlenbeck process,
\begin{equation}
\frac{df}{dt}=-\f{f}{\tau_{0}}+\eta_{2}~, 
\label{Langevin-2}
\end{equation}
where $\eta_{2}$ is another Gaussian white noise with mean zero and
correlation $\langle \eta_{2}(t)\eta_{2}(s)\rangle=2A\delta(t-s)$.
This system reaches a steady state and in the steady state the
external force has zero mean and covariance $\langle f(t) f(s)\rangle
=A\tau_0 \exp(-|t-s|/\tau_0)$.

The heat current flowing from the bath to the particle is the force
exerted by the bath times the velocity of the
particle \cite{Sekimoto:98}. Therefore, in a given time $\tau$, the
total amount of heat flow (in the unit of $K_B T$) is given by,
\begin{equation}
Q_{\tau}=\f{1}{k_{B}T}\int_{0}^{\tau}(-\gamma v+\eta_{1})v(t)dt~.
\label{Heat}
\end{equation}
On the other hand, the change in the internal energy of the particle in
this finite interval $\tau$ is given by
\begin{equation}
\Delta
U(\tau)=\f{1}{k_{B}T}\left[\f{1}{2}mv^{2}(\tau)-\f{1}{2}mv^{2}(0)\right]~. 
\label{Energy}
\end{equation}
Then the first law of the thermodynamics (conservation of energy)
gives $\Delta U(\tau)=W_{\tau}+Q_{\tau}$, where $W_{\tau}$ is the work
done on the particle by the external force, which is given by
\begin{equation}
W_{\tau}=\f{1}{k_{B}T}\int_{0}^{\tau}f(t)v(t)dt~.
\label{Work}
\end{equation}
This work is a stochastic quantity and our goal is to compute its PDF
$P(W_{\tau})$. 

It will prove convenient to introduce following two dimensionless parameters:
\begin{equation}
\theta=\f{\tau_{0}^{2}A}{D}, \quad\text{and} 
\quad\delta=\f{\tau_{0}}{\tau_{\gamma}}~. 
\label{theta-delta}
\end{equation}


\section{moment generating function}
\label{sec:MGF}

We begin by writing Eqs.~\eqref{Langevin-1} and \eqref{Langevin-2} in
the matrix form
\begin{equation}
\f{dU}{dt}=-AU+B\eta~,
\label{Langevin-matrix}
\end{equation}
where $U=(v,f)^{T}$ and $\eta=(\eta_{1},\eta_{2})^{T}$ are column
vectors, and $A$ and $B$ are $2\times2$ matrices given by
\begin{equation}
A=\begin{pmatrix}
1/\tau_{\gamma} & -1/m\\
0 & 1/\tau_{0}
\end{pmatrix},\quad
B=\begin{pmatrix}
1/m & 0\\
0 & 1
\end{pmatrix}.
\label{system-matrix}
\end{equation}
To compute the PDF of $W_{\tau}$ , we first consider its moment
generating function, constrained to fixed initial and final
configurations $U_{0}$ and $U$ respectively:
\begin{equation}
Z(\lambda,U,\tau|U_{0}) = \langle e^{-\lambda W_{\tau}} \delta[U-U(\tau)]
\rangle_{U_{0}}~,
\label{restricted GF}
\end{equation}
where the averaging is over the histories of the thermal noises
starting from the initial condition $U_{0}$. It is easy to show that
this restricted moment generating function satisfies the Fokker-Planck
equation
\begin{equation}
\f{\partial Z}{\partial \tau}=\mathcal{L}_{\lambda}Z~, 
\label{FP-eq}
\end{equation}
with the initial condition $Z(\lambda,U,0|U_{0})=\delta(U-U_{0})$. The
Fokker-Planck operator is given by
\begin{align}
\mathcal{L}_{\lambda}=\f{D}{m^{2}}\f{\partial^{2}}{\partial v^{2}}+\f{D \theta}{\tau_{0}^{2}}\f{\partial^{2}}{\partial f^{2}}+ \f{1}{\tau_{\gamma}} \f{\partial}{\partial v}v+\f{1}{\tau_{0}} \f{\partial}{\partial f}f \nn \\
-\f{f}{m} \f{\partial}{\partial v}-\f{\lambda \gamma}{D}fv~.
\label{FP-operator}
\end{align}
The solution of this equation can be formally expressed in the
eigenbases of the operator $\mathcal{L}_{\lambda}$ and the
large-$\tau$ behavior is dominated by the term containing the largest
eigenvalue. Thus, for large $\tau$ , one can write,
\begin{equation}
Z(\lambda,U,\tau|U_{0})=\chi(U_{0},\lambda)\Psi(U,\lambda)e^{\tau
\mu(\lambda)}+\dotsb, 
\label{characteristic.1}
\end{equation}
where $\mu(\lambda)$ is the largest eigenvalue,
$\mathcal{L}_{\lambda}\Psi(U,\lambda) =\mu(\lambda)\Psi(U,\lambda)$
and $\int dU \chi(U,\lambda) \Psi(U,\lambda)=1$.  Following the detail
calculation given in \aref{sec:Full-MGF}, we find that
\begin{subequations}
\begin{equation}
\mu(\lambda)=\f{1}{2\tau_{\gamma}}[1-\bar{\nu}(\lambda)]~, 
\label{mu}
\end{equation}
where 
\begin{equation}
\bar{\nu}(\lambda)=\f{1}{\delta}
\left[\sqrt{1+\delta^{2}+2\delta \nu(\lambda)}-1\right]~,  
\label{nu-bar}
\end{equation}
with 
\begin{equation}
\nu(\lambda)=\sqrt{1+4\theta\lambda(1-\lambda)}~. 
\label{nu}
\end{equation}
\end{subequations}
We note that $\mu(\lambda)$ obeys the so-called Gallavotti-Cohen
symmetry, 
\begin{math}
\mu(\lambda)=\mu(1-\lambda)~.
\end{math}

The moment generating function can be obtained by averaging the
restricted generating function over the initial variables $U_{0}$ with
respect to the steady state distribution $P_\mathrm{SS}(U_{0})$ and
integrating out the the final variables $U$,
\begin{equation}
Z(\lambda,\tau)=\int dU \int dU_{0} P_\mathrm{SS}(U_{0})Z(\lambda,U,\tau|U_{0})~,
\label{unrestricted-GF}
\end{equation}
where $P_\mathrm{SS}(U_{0})=\Psi(U_0,0)$.  This yields
\begin{equation}
Z(\lambda,\tau)=\langle e^{-\lambda
W_{\tau}}\rangle=g(\lambda)e^{\tau \mu(\lambda)}~+
\dotsb, 
\label{Z-asymptotic}
\end{equation}
where 
\begin{equation}
g(\lambda)=\int dU \int dU_{0}
\Psi(U_0,0)\chi(U_{0},\lambda)\Psi(U,\lambda)~. 
\label{glambda-1}
\end{equation}
The full forms of $\Psi(U,\lambda)$ and $\chi(U_{0},\lambda)$ are
given by \eref{psi-chi}. Using these we find the $g(\lambda)$ as given
by Eqs.~\eqref{glambda-2} and \eqref{glambda-3}
in \aref{sec:Full-MGF}.


\section{Probability Distribution function}
\label{sec:PDF}

The PDF $P(W_{\tau})$ is related to the moment generating function
$Z(\lambda,\tau)$ as
\begin{equation}
P(W_{\tau})=\f{1}{2\pi
i}\int_{-i\infty}^{+i\infty}Z(\lambda,\tau)e^{\lambda
W_{\tau}}d\lambda~,
\end{equation}
where the integration is done in the complex $\lambda$
plane. Inserting the large $\tau$ form of $Z(\lambda,\tau)$ given
by \eref{Z-asymptotic}, we obtain
\begin{equation}
P(W_{\tau}=w\tau /\tau_{\gamma})\approx \f{1}{2\pi i}\int_{-i\infty}^{+i\infty}g(\lambda)e^{(\tau/\tau_{\gamma})\, f_{w}(\lambda)}d\lambda~,
\label{P(W)}
\end{equation}
where
\begin{equation}
f_{w}(\lambda)=\f{1}{2}[1-\bar{\nu}(\lambda)] +\lambda
w~.
\label{f_{w}}
\end{equation}
In the large $\tau$ limit, we can use the saddle point approximation,
in which one chooses the contour of integration along the steepest
descent path through the saddle point $\lambda^{*}$. The saddle point
can be obtained solving the equation,
\begin{equation} 
f^{\prime}_{w}(\lambda^{*})=0~, 
\end{equation}
or equivalently,
\begin{equation}
\bar{\nu}^{\prime}(\lambda^{*})=2w~.
\label{saddle-eqn-1}
\end{equation}
The above equation yields 
\begin{equation}
\theta(1-2\lambda^{*})=w\nu(\lambda^{*})
\sqrt{1+\delta^{2}+2\delta\nu(\lambda^{*})}~.
\label{saddle-eqn-2}
\end{equation}
Since $\theta$, $\delta$ and $\nu(\lambda)$ are always positive, it is
clear that sign$(1-2\lambda^{*})$=sign$(w)$. The above equation
can be simplified to the cubic form
\begin{equation}
\nu^{3}(\lambda^{*})+a\nu^{2}(\lambda^{*})-b=0~, 
\label{saddle-eqn-3}
\end{equation}
where
\begin{subequations}
\label{a-b}
\begin{align}
a&=\f{\theta+(1+\delta^{2})w^{2}}{2\delta w^{2}}~,   \\
b&=\f{\theta+\theta^{2}}{2\delta w^{2}}~.
\end{align}
\end{subequations}
We observe that one of the roots of the cubic equation for
$\nu(\lambda^{*})$ is real while the other two are
complex. \Eref{saddle-eqn-2} suggests the root to be real, and it is
given by
\begin{subequations}
\label{nu-lambda-star}
\begin{align}
\nu(\lambda^{*})=-\f{a}{3}\Bigl[1&-
\bigl(1+2~k+3\sqrt{3~l~k}\bigr)^{-1/3}\nn \\
&-\bigl(1+2~k+3\sqrt{3~l~k}\bigr)^{1/3}\Bigr],
\end{align}
where $l=b/a^{3}$ and $k=(27/4)\, l-1$.
Note that $l > 0$. Therefore, $\nu(\lambda^*)$ is evidently real for
$k>0$.  On the other hand, when $k<0$, it can be simplified to the
evidently real form
\begin{align}
\nu(\lambda^{*})=&-\f{a}{3}\bigl[1-2~\cos\,(\phi/3)\bigr]~,
\end{align}
\end{subequations}
where $\phi=\tan^{-1}\bigl[3\sqrt{3l|k|}\big/(1+2k)\bigr] \in
[0,\pi]$.  

In the limit $w\to \pm \infty$, from \eref{a-b} we have, $a\to
(1+\delta^2)/(2 \delta)$ and $b\to 0$. Therefore, $l\to 0$ and $k\to
-1$, giving $\phi\to \pi$. This yields, $\nu(\lambda^*)\to 0$. On the
other hand, for $w\to 0$, we have, $a\sim \theta/(2\delta w^2)$. Using
this we find that $\nu(\lambda^*) \to \sqrt{1+\theta}$. It is also
evident as \eref{saddle-eqn-2} gives $\lambda^*=1/2$ for $w=0$, and
then, from \eref{nu} we get $\nu(1/2)=\sqrt{1+\theta}$.

\begin{figure}
\includegraphics[width=.9\hsize]{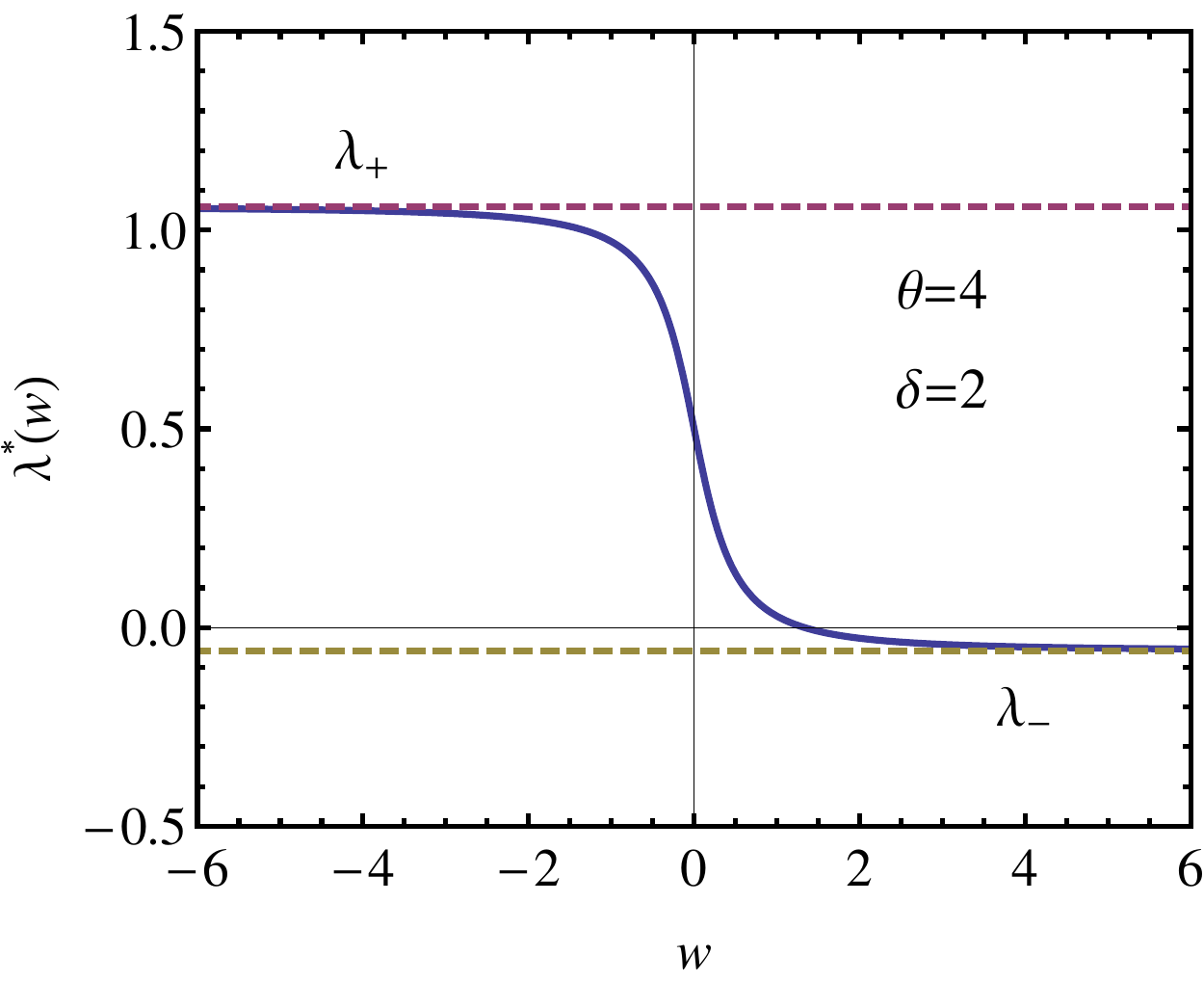}
\caption{\label{saddle-LDF}(Color online) The behavior of 
$\lambda^{*}$ is shown (solid line) as a function of $w$, for a set of
parameters $\theta=4,~\delta=2$, which merges to $\lambda_{\pm}$
(dashed lines) as $w \to \mp \infty$.}
\end{figure}

Now using \eref{saddle-eqn-2}, the saddle point $\lambda^*(w)$ can be
expressed in terms of $\nu(\lambda^*)$. Therefore, the function
$f_w(\lambda)$ at the saddle-point $\lambda^*$, can be expressed in
terms of $\nu(\lambda^*)$, and is given by
\begin{align}
&h_s(w):=f_{w}(\lambda^{*})\nn\\
&=\f{1}{2}\left[\frac{1}{\delta}+1+w\right]
-\f{1}{2}\left[\f{1}{\delta}+\f{w^{2}}{\theta}\nu(\lambda^{*})\right]
\sqrt{1+\delta^{2}+2\delta \nu(\lambda^{*})}~.
\label{LDF-saddle}
\end{align}
To find the region in which $\lambda^*$ lies, it is useful to
express $\nu(\lambda)$ in the form
\begin{equation}
\nu(\lambda)=\sqrt{4 \theta
(\lambda_{+}-\lambda)(\lambda-\lambda_{-})}~, 
\end{equation}
where 
\begin{equation}
\lambda_{\pm}=\f{1}{2}\left[1\pm\sqrt{1+\theta^{-1}}\right]~. 
\label{lambda-pm}
\end{equation}
Clearly, $\nu(\lambda)$ has two branch points on the real-$\lambda$
line at $\lambda_{\pm}$. Moreover, it is real and positive in the
(real) interval $\lambda \in (\lambda_{-},\lambda_{+})$. Since,
$\lambda_+ - \lambda_-= \sqrt{1+\theta^{-1}}$, as
$\lambda\to \lambda_\pm$, we have $\nu(\lambda)\to 2 [\theta
(1+\theta)]^{1/4} |\lambda-\lambda_\pm|^{1/2}$. Therefore,
from \eref{saddle-eqn-2} we get
\begin{equation}
w \to \mp\, \frac{[\theta (1+\theta)]^{1/4}}{2\sqrt{1+\delta^2}}
|\lambda^*-\lambda_\pm|^{-1/2}, \quad \text{as}~\lambda^*\to \lambda_\pm~. 
\end{equation}
In other words, $\lambda^{*}(w)$ merges to $\lambda_{\pm}$ as one
takes the limit $w\rightarrow \mp \infty$. This also agrees with the
observation that $\nu(\lambda^*)\to 0$ as $|w|\to \infty$. For any
finite $w$ the saddle point $\lambda^{*} \in
(\lambda_{-},\lambda_{+})$.  In \fref{saddle-LDF} we plot the saddle
point $\lambda^{*}$ as a function of $w$ using \eref{saddle-eqn-2}.
 
Now, if $g(\lambda)$ is analytic in the range $\lambda \in
(0,\lambda^{*})$, we can deform the contour along the path of the
steepest descent through the saddle point, and obtain $P(W_{\tau})$
using the usual saddle point method . However, more sophistication is
needed when $g(\lambda)$ contains singularities. Therefore it is
essential to analyze $g(\lambda)$ for possible singularities.


\begin{figure}
\includegraphics[width=.85\hsize]{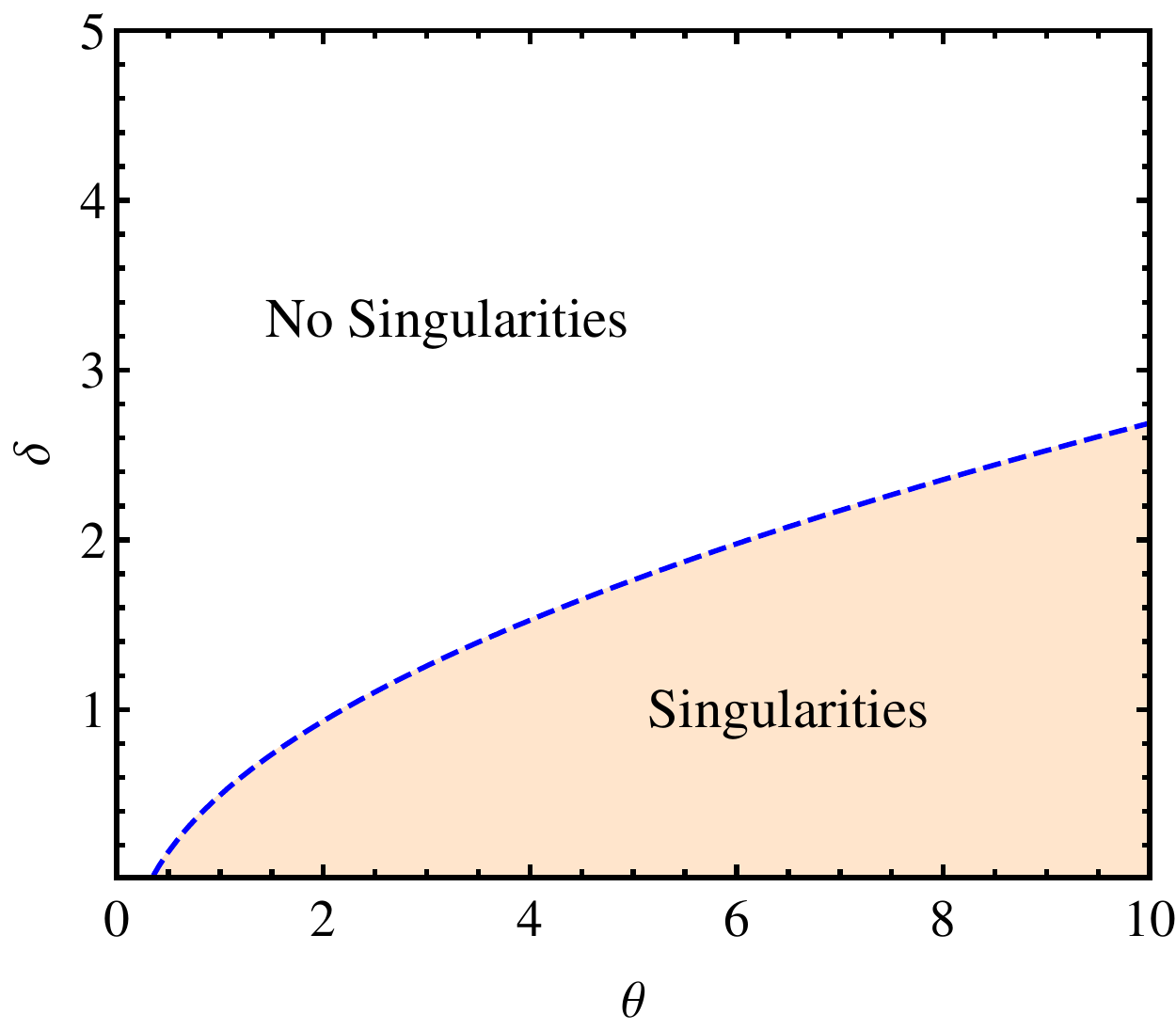}
\caption{\label{phase-space}(Color online) This plot depicts the
analytic properties of $g(\lambda)$. In the shaded region of the
$(\theta,\delta)$ plane, $g(\lambda)$ possesses a singularity, where
$f_{2}(\lambda_+,\theta,\delta)<0$. On the other hand, in the unshaded
region $g(\lambda)$ does not have any singularities, where
$f_{2}(\lambda_+,\theta,\delta)>0$. These two domains are separated by
the boundary given by the equation
$f_{2}(\lambda_{+},\theta,\delta)=0$.}
\end{figure}

We first recall $g(\lambda)$ from \eref{glambda-2}
and \eref{glambda-3},
\begin{equation}
g(\lambda)=\bigl[f_{1}(\lambda,\theta,\delta)\bigr]^{-1/2} 
\bigl[f_{2}(\lambda,\theta,\delta)\bigr]^{-1/2}~. 
\end{equation}
Following \aref{sec:Full-MGF}, we also recall that
$f_{1}(\lambda,\theta,\delta)$ does not change its sign and always
stays positive in the region $[\lambda_{-},\lambda_{+}]$.  This is not
the case for $f_{2}(\lambda,\theta,\delta)$.  While
$f_2(\lambda,\theta,\delta) >0$ for $\lambda_- \le \lambda \le 0$, in
some region in the $(\theta, \delta)$ space,
$f_{2}(\lambda_+,\theta,\delta) <0$. Therefore, in that
$(\theta, \delta)$ region, $f_{2}(\lambda,\theta,\delta)$ must have a
zero at some intermediate $\lambda=\lambda_0 >0$, which gives rise to
a branch-point singularity in $g(\lambda)$. \Fref{phase-space} shows
parameter region in which $g(\lambda)$ possesses a singularity. The
phase boundary between the region which $g(\lambda)$ has a singularity
and the singularity-free region is given by the equation
$f_{2}(\lambda_{+},\theta,\delta)=0$. In the limit $\delta\to 0$ we
get $\theta\to 1/3$.

\subsection{Case of no singularities}
\label{subsec:analytical}

\begin{figure}
\includegraphics[width=.95\hsize]{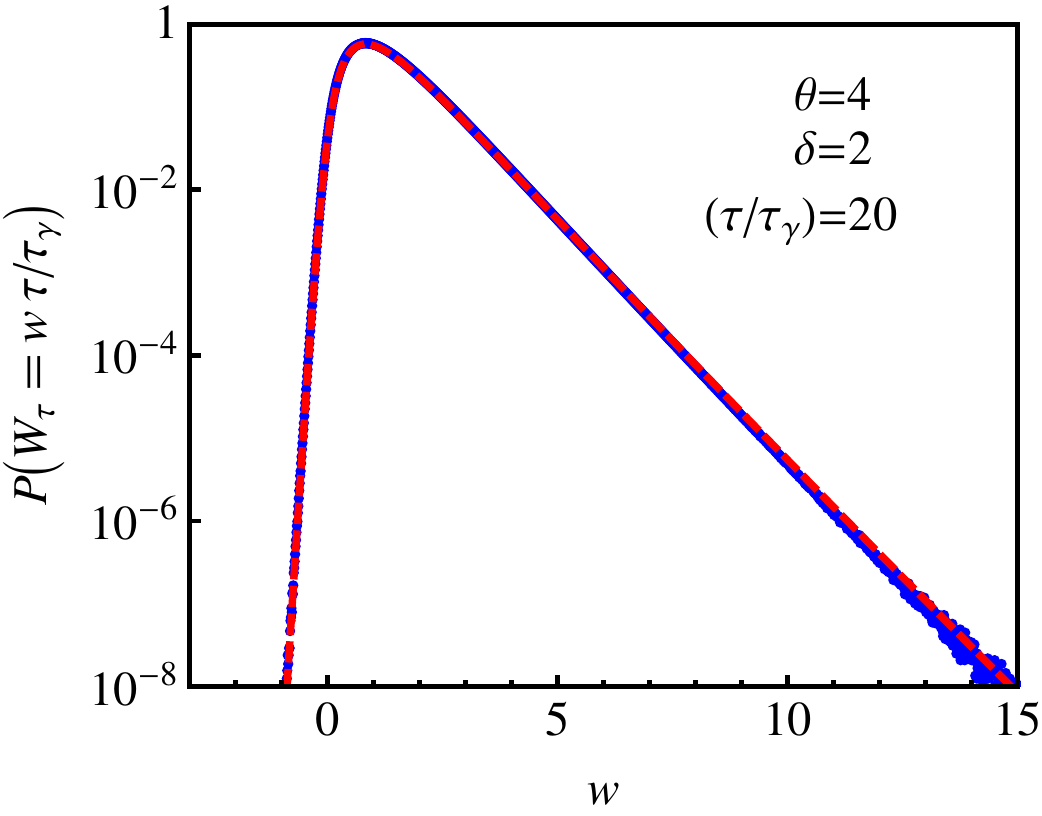}
\caption{\label{saddle-PDF-plot}(Color online) The (red) dashed line
plots the analytical result of $P(W_{\tau})$ against the scaled
variable $w=W_{\tau}/(\tau/\tau_{\gamma})$, while the (blue) points
are numerical simulation results.}
\end{figure}

In the singularity free region (\fref{phase-space}), the asymptotic
PDF of the work done is obtained using the standard saddle point
method, which gives
\begin{equation}
P(W_{\tau}=w\tau /\tau_{\gamma})\approx\f{g(\lambda^{*})e^{\f{\tau}{\tau_{\gamma}} h_{s}(w)}}{\sqrt{2 \pi \f{\tau}{\tau_{\gamma}}f_{w}^{\prime\prime}(\lambda^{*})}}~,
\label{PDF-saddle}
\end{equation}
where $h_{s}(w)$ is given by \eref{LDF-saddle} and 
\begin{equation}
f_{w}^{\prime\prime}(\lambda^{*})=-\f{\bar{\nu}^{\prime \prime}(\lambda^{*})}{2}
=\f{2}{\nu(\lambda^{*})}~\f{\theta+w^{2}[1+\delta^{2}+3\delta \nu(\lambda^{*})]}{[1+\delta^{2}+2\delta \nu(\lambda^{*})]^{1/2}},
\label{fwdoubleprime}
\end{equation}
which is expressed in terms of $w$ and $\nu(\lambda^{*})$ given
by \eref{nu-lambda-star}.
\fref{saddle-PDF-plot} shows a very good agreement between the
analytic result given by \eref{PDF-saddle} and numerical
simulations.

\subsection{Case of a singularity}
\label{subsec:singularity}

For a given value of $\delta$ and $\theta$, the location of the branch
point $\lambda_{0}$ is fixed between the origin and $\lambda_{+}$. On
the other hand, the saddle point $\lambda^{*}$ increases monotonically
along the real-$\lambda$ line from $\lambda_{-}$ to $\lambda_{+}$ as
$w$ decreases from $+\infty$ to $-\infty$. For sufficiently large $w$,
the saddle point lies in the interval $(\lambda_{-},\lambda_{0})$ and
therefore, the contour of integration can be deformed into the
steepest descent path, which passes through the saddle point, without
touching $\lambda_{0}$. However, as $w$ decreases, the saddle point
hits the branch point at some specific value $w=w^{*}$ given by
\begin{equation}
\lambda^{*}(w^{*})=\lambda_{0}~.
\label{lambda-singularity}
\end{equation}

For $w<w^{*}$, the steepest descent contour wraps around the branch
cut between $\lambda_{0}$ and $\lambda^{*}$. We here present the
results for both regimes $w<w^{*}$ and $w>w^{*}$ respectively,
applying the method developed in \cite{Arnab:13}.

\subsubsection{$ w>w^{*} $}

For $w>w^{*}$, the contour is deformed through the saddle point
without touching the singularity and we obtain
\bea
P(W_{\tau}=w\tau /\tau_{\gamma})\approx\f{g(\lambda^{*})e^{\f{\tau}{\tau_{\gamma}} h_{s}(w)}}{\sqrt{2 \pi \f{\tau}{\tau_{\gamma}}f_{w}^{\prime\prime}(\lambda^{*})}}R_{1}\biggl(\sqrt{\f{\tau}{\tau_{\gamma}}[h_{0}(w)-h_{s}(w)]}\biggr), \nn \\
\label{PDF-singularity-1}
\eea
where $f_{w}^{\prime\prime}(\lambda^{*})$
is given by \eref{fwdoubleprime} and the function $R_{1}(z)$ is given by
\begin{equation}
R_{1}(z):=\f{z}{\sqrt{\pi}}e^{z^{2}/2}K_{1/4}(z^{2}/2)~,
\label{R1}
\end{equation}
with $K_{1/4}(z)$ being the modified Bessel function of the second kind. 
\subsubsection{$ w<w^{*} $}
\begin{figure}
\includegraphics[width=.95\hsize]{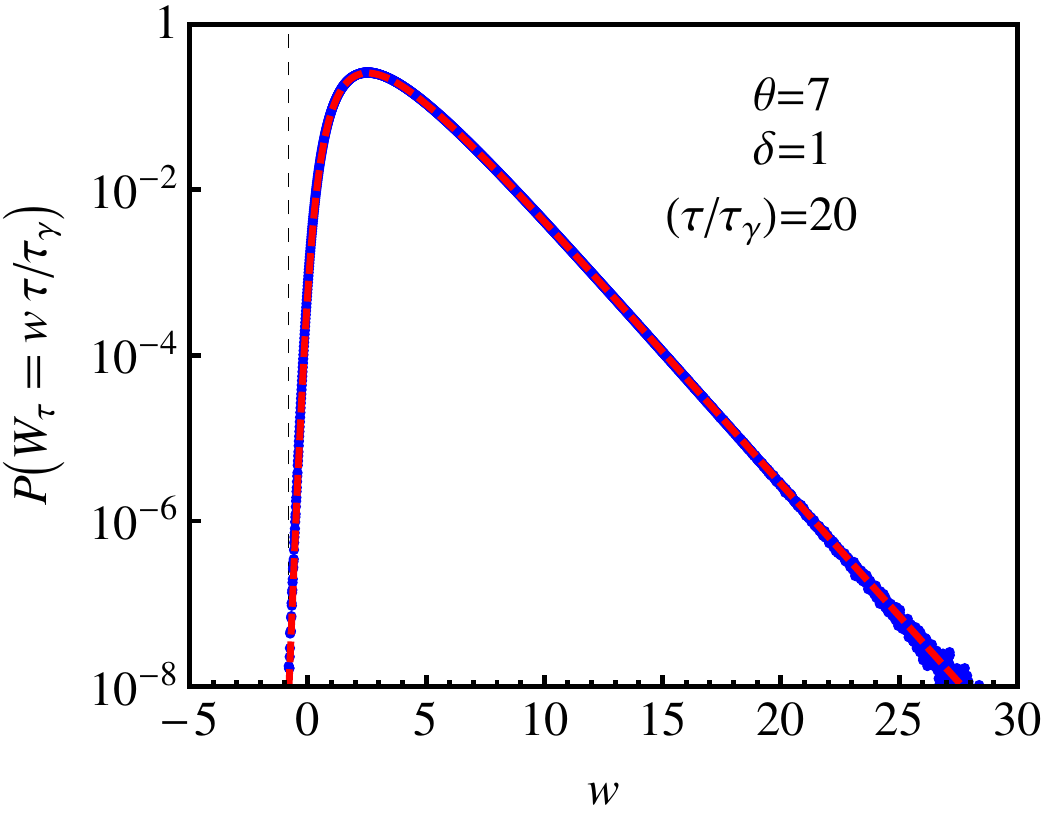}
\caption{\label{singularity-PDF-plot} (Color online) The (red) dashed
line plots the analytical result for $P(W_{\tau})$, while the (blue)
points are numerical simulation results. The vertical dashed line
marks the position of the singularity $w^{*}=-0.801661...$ for the
values of $\theta=7,~\delta=1$.}
\end{figure}

For $w<w^{*}$, the contribution comes from both the branch point and the saddle point i.e.
\begin{equation}
P(W_{\tau})\approx P_{B}(W_{\tau})+P_{S}(W_{\tau})~,
\end{equation}
where the branch point contribution is 
\begin{equation}
P_{B}(W_{\tau}=w\tau /\tau_{\gamma})\approx \f{\tilde{g}(\lambda_{0})e^{\f{\tau}{\tau_{\gamma}} h_{0}(w)}}{\sqrt{\pi \f{\tau}{\tau_{\gamma}}|f_{w}^{\prime}(\lambda_{0})}|}R_{2}\biggl(\sqrt{\f{\tau}{\tau_{\gamma}}[h_{0}(w)-h_{s}(w)]}\biggr), 
\label{PDF-singularity-2}
\end{equation}
where
\begin{align}
\label{LDF-singularity}
h_{0}(w)&:=f_{w}(\lambda_{0})=\f{1}{2}[1-\bar{\nu}(\lambda_{0})]+\lambda_{0}w~,\\
f_{w}^{\prime}(\lambda_{0})&=-\frac{\bar{\nu}^{\prime}(\lambda_{0})}{2}+w, \\
\label{gtilde}
\tilde{g}(\lambda_{0})&=\lim_{\lambda \to \lambda_{0}}|\sqrt{\lambda-\lambda_{0}}~g(\lambda)|~,\\
\intertext{and}
\label{R2}
R_{2}(z)&=\sqrt{\f{2z}{\pi}}\int_{0}^{z}\f{1}{\sqrt{u}}e^{-2zu+u^{2}}~du~.
\end{align}
The contribution coming from the saddle point is given by 
\begin{equation}
P_{S}(W_{\tau}=w\tau /\tau_{\gamma})\approx \f{|g(\lambda^{*})|e^{\f{\tau}{\tau_{\gamma}} h_{s}(w)}}{\sqrt{2\pi \f{\tau}{\tau_{\gamma}}|f_{w}^{\prime \prime}(\lambda^{*})}|}R_{4}\biggl(\sqrt{\f{\tau}{\tau_{\gamma}}[h_{0}(w)-h_{s}(w)]}\biggr), 
\label{PDF-singularity-3}
\end{equation}
where the function $R_{4}(z)$ is given by
\bea
R_{4}(z)=\sqrt{\f{\pi}{2}}ze^{z^{2}/2}\biggl[I_{-1/4}(z^{2}/2)+I_{1/4}(z^{2}/2)\biggr] \nn \\
-\f{4z}{\pi}{}_{2}F_{2}(1/2,1;3/4,5/4;z^{2})~,
\label{R4}
\eea
and $I_{\pm 1/4}(z)$ are modified Bessel functions of the first kind and ${}_{2}F_{2}(a_{1},a_{2};b_{1},b_{2};z)$ is the generalized hypergeometric function.
We again find a very good agreement between the analytical results and numerical simulations \fref{singularity-PDF-plot}.

In the following we analyze the $\delta=0$ case, which becomes a
special case of the problem of a single Brownian particle connected
with two heat baths at different temperature studied by
Visco\cite{Visco:06}. Here, we obtain the PDF.


\subsection{$\delta=0$}

\begin{figure}
\includegraphics[width=.93\hsize]{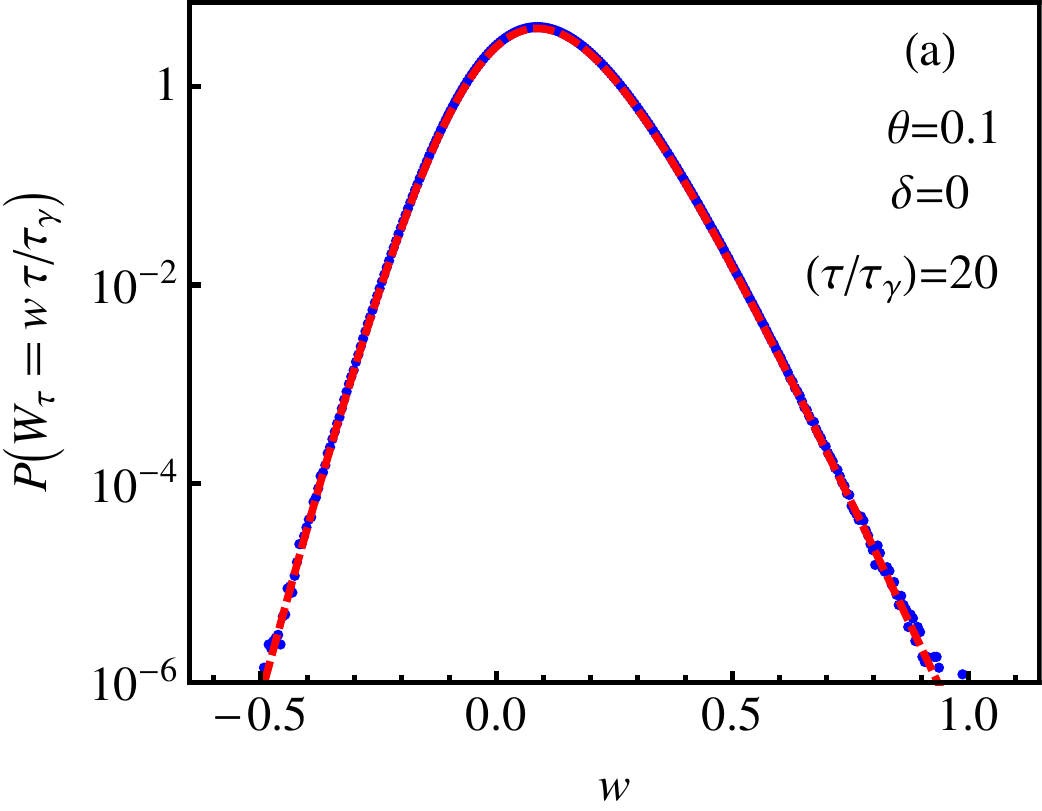}
\includegraphics[width=.95\hsize]{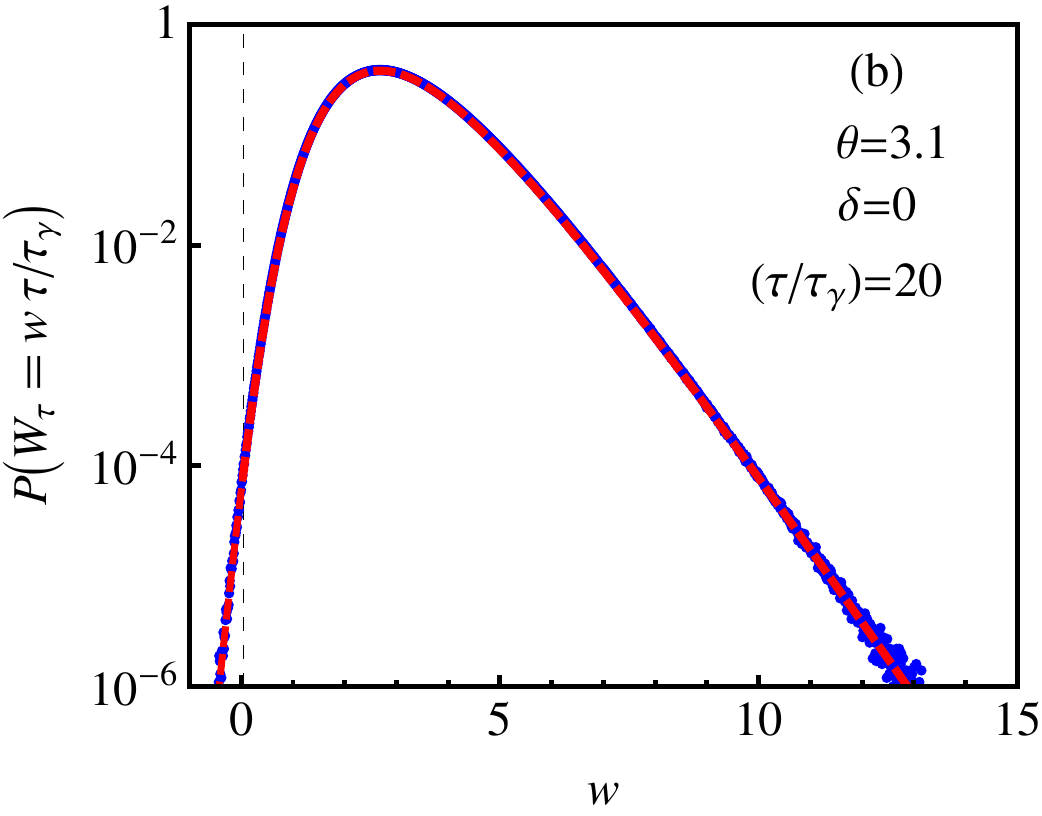}
\caption{\label{saddle-PDF-plot-delta-zero}(Color online) The (red)
dashed lines plot analytical results for $P(W_{\tau})$, while the
(blue) points are numerical simulation results, for the $\delta=0$
case. The vertical dashed line in (b) marks the position of the
singularity which is $w^{*}=0.037...$ in this case. }
\end{figure}

 We first note that, $g(\lambda)$ takes a simple form in the limit
$\delta\to 0$, given by,
\begin{equation}
g(\lambda)=\f{\sqrt{2 \nu}}{\sqrt{\nu+1+2 \lambda \theta}} \f{\sqrt{2}}{\sqrt{\nu+1-2 \lambda \theta}}~.
\end{equation}
It is easy to show~\cite{Sabhapandit:11-12} that $g(\lambda)$ is
completely analytic for $\theta \leq 1/3$, and the PDF is obtained
using the saddle point method as,
\begin{equation}
P(W_{\tau}=w\tau /\tau_{\gamma})\approx\f{g(\lambda^{*})e^{\f{\tau}{\tau_{\gamma}} h_{s}(w)}}{\sqrt{2 \pi \f{\tau}{\tau_{\gamma}}f_{w}^{\prime\prime}(\lambda^{*})}}~,
\end{equation}
where the second derivative of $f_{w}(\lambda)$ along the
real-$\lambda$ axis at $\lambda^{*}$ is given
by~\cite{Sabhapandit:11-12},
\begin{equation}
f_{w}^{\prime\prime}(\lambda^{*})=\f{2(w^{2}+\theta)^{3/2}}{\sqrt{\theta(1+\theta)}}~,
\end{equation}
and 
\begin{equation}
h_{s}(w):=f_{w}(\lambda^{*})
=\f{1}{2}\biggl[1+w-\sqrt{w^{2}+\theta}\sqrt{1+\f{1}{\theta}}\biggr]~.
\end{equation}
On the other hand, if $\theta > 1/3$, it is easy to show that
$g(\lambda)$ picks up a branch point singularity at
$\lambda=\lambda_{0}=2/(1+\theta)$, which corresponds
to~\cite{Sabhapandit:11-12},
\begin{equation}
w^{*}=\f{\theta(\theta-3)}{3\theta-1}~.
\label{wstar}
\end{equation}
Then one needs to perform a contour integration avoiding the branch
cut as mentioned in the last section.  For $w>w^{*}$, using the same
prescription \cite{Arnab:13}, we find the PDF as
\begin{equation}
P(W_{\tau}=w\tau /\tau_{\gamma})\approx\f{g(\lambda^{*})e^{\f{\tau}{\tau_{\gamma}} h_{s}(w)}}{\sqrt{2 \pi \f{\tau}{\tau_{\gamma}}f_{w}^{\prime\prime}(\lambda^{*})}}R_{1}\biggl(\sqrt{\f{\tau}{\tau_{\gamma}}[h_{0}(w)-h_{s}(w)]}\biggr), 
\end{equation}
where 
\begin{equation}
h_{0}(w):=f_{w}(\lambda_{0})=\f{1-\theta}{1+\theta}+\f{2~w}{1+\theta}~.
\end{equation}
For $w<w^{*}$, the contribution to the PDF comes both from the saddle
and the branch point.
\begin{equation}
P(W_{\tau})\approx P_{B}(W_{\tau})+P_{S}(W_{\tau})~,
\end{equation}
where the branch point contribution is 
\bea
P_{B}(W_{\tau}=w\tau /\tau_{\gamma})\approx \f{\tilde{g}(\lambda_{0})e^{\f{\tau}{\tau_{\gamma}} h_{0}(w)}}{\sqrt{\pi \f{\tau}{\tau_{\gamma}}|f_{w}^{\prime}(\lambda_{0})}|}R_{2}\biggl(\sqrt{\f{\tau}{\tau_{\gamma}}[h_{0}(w)-h_{s}(w)]}\biggr), \nn \\
\label{PDF-singularity-2}
\eea
where 
\bea
\tilde{g}(\lambda_{0})&=&\f{3\theta-1}{2\theta\sqrt{2(1+\theta)}}~, \nn \\
f_{w}^{\prime}(\lambda_{0})&=&w-w^{*}~,
\eea
and the function $R_{2}(z)$ is given by \eref{R2}.
The contribution coming from the saddle point is given by 
\bea
P_{S}(W_{\tau}=w\tau /\tau_{\gamma})\approx \f{|g(\lambda^{*})|e^{\f{\tau}{\tau_{\gamma}} h_{s}(w)}}{\sqrt{2\pi \f{\tau}{\tau_{\gamma}}|f_{w}^{\prime \prime}(\lambda^{*})}|}R_{4}\biggl(\sqrt{\f{\tau}{\tau_{\gamma}}[h_{0}(w)-h_{s}(w)]}\biggr), \nn \\
\label{PDF-singularity-3}
\eea
where the function $R_{4}(z)$ is given by \eref{R4}. 
\Fref{saddle-PDF-plot-delta-zero} compares the analytical results with
the numerical simulations.

\section{Large deviation function and the fluctuation theorems}
\label{sec:FT}

The LDF, associated with the PDF, is defined as
\begin{equation}
h(w)=\lim_{(\tau/\tau_\gamma) \to \infty}~\f{1}{(\tau/\tau_\gamma)}~
\ln~P(W_{\tau}=w \tau/\tau_{\gamma})~. 
\end{equation}
Due to the large deviation form of the PDF,
$P(W_{\tau}=w \tau/\tau_{\gamma})\sim e^{(\tau/\tau_{\gamma})\,
h(w)}$, the FT given by \eref{Fluctuation Theorem}, is equivalent to
the following symmetry relation of the LDF:
\begin{equation}
h(w)-h(-w)=w~.
\label{symmetry-1}
\end{equation}

Now, in the parameter region where $g(\lambda)$ is analytic
[see \fref{phase-space}], the LDF is given by $h(w)=h_s(w)$. In this
case, it is clear from \eref{LDF-saddle} that the above symmetry
relation \eqref{symmetry-1} holds, as $\nu(\lambda^*)$ is an even
function in $w$.

On the other hand, in the parameter region where $g(\lambda)$ has a
singularity, the LDF is given by
\begin{equation}
h(w)=   
\begin{cases}
    h_{s}(w) & \text{for}~w~>~w^{*}~,\\
    h_{0}(w) & \text{for}~w~<~w^{*}~.
\end{cases}
\end{equation}
Therefore, it is evident that if $w^{*}<0$, the symmetry
relation \eqref{symmetry-1} holds only in the specific range
$w^{*}<w<-w^{*}$. Otherwise, it fails to satisfy.  Nevertheless, even
for $w>w^*$, one still gets a linear relation $h(w)-h(-w)=2\lambda_0 w$,
in the range $w\in (-w^*,w^*)$.

\section{Summary}
\label{sec:Summary}

In this paper, we have discussed an underdamped Brownian particle
driven by an external correlated stochastic force, modeled by an
Ornstein-Uhlenbeck process. We have studied the probability density
function (PDF) of the work done $W_\tau$ on the particle by the
external random force, in a given time $\tau$.  The behavior can be
characterized in terms of two dimensionless parameters, namely, (i)
$\theta$, that gives the relative strength between the external random
force and the thermal noise, and (ii) $\delta$, that characterizes the
ratio between the the viscous relaxation time and the correlation time
of the external force. In the large $\tau$ limit, we have obtained the
moment generating function (MGF) in the form, $\langle e^{-\lambda
W_{\tau}}\rangle \sim g(\lambda) e^{\tau \mu(\lambda)}$. While
$\mu(\lambda)$ is analytic in the relevant region of $\lambda$ (where
the saddle point lies), the prefactor $g(\lambda)$ shows analytical as
well as singular behavior in different parts of the parameter space
spanned by $(\theta,\delta)$. We have obtained the PDF in both
analytic and non-analytic regions of $(\theta,~\delta)$ space, by
carefully inverting the MGF.  The entire analytical results have been
supported by numerical simulations.  In the limit $\delta \to 0$, our
model becomes a special case of a problem of a single Brownian
particle coupled to two distinct reservoirs, first proposed by Derrida
and Brunet \cite{Brunet:05} and later studied by
Visco \cite{Visco:06}.

We have also looked at the validity of the fluctuation theorem (FT)
for work, in terms of the symmetry properties of the large deviation
function. We have found that in the $(\theta, \delta)$ region where
$g(\lambda)$ is analytic, the FT is satisfied. On the other hand, in
the non-analytic region, the symmetry of the large deviation function
breaks down.  In particular, the PDF picks up an exponential tail
characterized by the singularity and this leads to the violation of
the steady state fluctuation theorems.

Finally, we have provided a non-trivial example where the exact LDF as
well as the complete asymptotic form of the PDF of the work can be
computed.


\begin{acknowledgements}
The authors thank the Galileo Galilei Institute for Theoretical
Physics, Florence, Italy for the hospitality and the INFN for partial
support during the completion of this work. SS acknowledges the
support of the Indo-French Centre for the Promotion of Advanced
Research under Project 4604-3.
\end{acknowledgements}


\appendix
\section{Detailed calculation of the MGF}
\label{sec:Full-MGF}

We recall \eref{Langevin-matrix} and \eref{system-matrix}
\begin{equation}
\f{dU}{dt}=-AU+B\eta~,
\end{equation}
where $U=(v,f)^{T}$ and $\eta=(\eta_{1},\eta_{2})^{T}$ are column
vectors and $A$, $B$ are $2\times2$ matrices given by
\begin{equation}
A=\begin{pmatrix}
1/\tau_{\gamma} & -1/m\\
0 & 1/\tau_{0}
\end{pmatrix},\quad
B=\begin{pmatrix}
1/m & 0\\
0 & 1~
\end{pmatrix}.
\end{equation}
The expression for $W_{\tau}$ can then be expressed in terms of these matrices
\begin{equation}
W_{\tau}=\f{\gamma}{2D}\int_{0}^{\tau}dt~U^{T}A_{1}U~,
\end{equation}
where $A_{1}$ is a real symmetric matrix
\begin{equation}
A_{1}=\begin{pmatrix}
0 & 1\\
1 & 0~
\end{pmatrix}.
\end{equation}
Using the integral representation of the delta-function, we rewrite
the moment generating function
\begin{equation}
Z(\lambda,U,\tau|U_{0})=\int \frac{d^{2}\sigma}{(2\pi)^{2}} e^{i\sigma^{T}U} \langle e^{-\lambda W_{\tau}-i\sigma^{T}U(\tau)}\rangle_{U,U_{0}}~.
\label{R-GF}
\end{equation}
Now, we proceed by defining the finite time Fourier transforms and
inverses as follows:
\begin{subequations}
\begin{align}
\lbrack \tilde{U}(\omega_{n}),\tilde{\eta}(\omega_{n}) \rbrack&= \frac{1}{\tau} \int_{0}^{\tau} dt \lbrack U(t),\eta(t) \rbrack \exp(-i \omega_{n} t), \\
\lbrack U(t),\eta(t) \rbrack&= \sum_{n=-\infty}^{\infty} \lbrack \tilde{U}(\omega_{n}),\tilde{\eta}(\omega_{n}) \rbrack \exp(i \omega_{n} t),
\end{align}
\end{subequations}
with $\omega_{n}=2\pi n/\tau$.

In the frequency domain, the Gaussian noise configurations denoted by
$\{\eta(t):0<t<\tau\}$ can be well described by the infinite
sequence
$\{\tilde{\eta}(\omega_{n}):n=-\infty,...,-1,0,+1,...,\infty\}$ of
Gaussian random variables having the following correlations
\begin{equation}
\langle \tilde{\eta}(\omega) \tilde{\eta}^{T}(\omega^{\prime})\rangle=\frac{2D}{\tau} \delta(\omega+\omega^{\prime})~{\rm diag}(1,\theta/\tau_{0}^{2})~.
\end{equation}
The Fourier transform of $U(t)$ is then straightforward and henceforth
the expression for $W_{\tau}$  becomes 
\begin{equation}
\tilde{U}=GB\tilde{\eta}-\f{1}{\tau}G\Delta U\\
W_{\tau}=\f{\gamma \tau}{2D}\sum_{n=-\infty}^{\infty}\tilde{U}^{T}(\omega_{n})A_{1}\tilde{U}^{*}(\omega_{n})~,
\label{work-1}
\end{equation}
where $G(\omega)=(i\omega I+A)^{-1}$ and $\Delta U=U(\tau)-U(0)$, with
$I$ being the identity matrix. The elements of $G$ are
$G_{11}=\tau_{\gamma}(i\omega\tau_{\gamma}+1)^{-1},~G_{22}=
\tau_{0}(i\omega\tau_{0}+1)^{-1},~G_{12}=G_{11}G_{22}/m,~G_{21}=0$. Substituting
$\tilde{U}$ from the above expression in $W_{\tau}$ and grouping the
negative indices into their positive counterparts, we obtain 
\begin{align}
W_{\tau}=\f{\gamma \tau}{2D}\biggl[\tilde{\eta}^{T}_{0}(BG^{T}_{0}A_{1}G_{0}B)\tilde{\eta}_{0}-\f{2}{\tau}\Delta
U^{T}(G^{T}_{0}A_{1}G_{0}B)\tilde{\eta}_{0}\notag \\ 
+\f{1}{\tau^{2}}\Delta
U^{T}(G^{T}_{0}A_{1}G_{0})\Delta U \biggr]\notag\\
+\f{\gamma \tau}{D}\sum_{n=1}^{\infty}\biggl[\tilde{\eta}^{T}(BG^{T}A_{1}G^{*}B)\tilde{\eta}^{*} 
-\f{1}{\tau}\Delta U^{T}(G^{T}A_{1}G^{*}B)\tilde{\eta}^{*}\notag \\
-\f{1}{\tau}\tilde{\eta}^{T}(BG^{T}A_{1}G^{*})\Delta U
+\f{1}{\tau^{2}}\Delta U^{T}(G^{T}A_{1}G^{*})\Delta U \biggr]~,
\label{work-2}
\end{align}
where
$G_{0}=G(\omega=0)=A^{-1},~\tilde{\eta_{0}}=\tilde{\eta}(0)$. The
finite time Fourier series can be written for $U(\tau)$ as well
\begin{align}
U(\tau)&=\lim_{\epsilon \rightarrow0}\sum_{n=-\infty}^{\infty} \tilde{U}(\omega_{n}) e^{-i\omega_{n}\epsilon}\notag \\
&=\lim_{\epsilon \rightarrow0}\sum_{n=-\infty}^{\infty}(GB\tilde{\eta}-\f{1}{\tau}G\Delta U)e^{-i\omega_{n}\epsilon}\notag\\
&=\lim_{\epsilon \rightarrow0}\sum_{n=-\infty}^{\infty}(GB\tilde{\eta})e^{-i\omega_{n}\epsilon}~,
\end{align}
where we observe that $\tau^{-1} \sum_{n} G(\omega_{n})
e^{-i\omega_{n}\epsilon}=0$ for large $\tau$. This is because while
converting the summation into an integral we note that all the poles
of $G(\omega)$ lie in the upper half plane. In other words, the
function $G(\omega)$ is analytic in the lower half. Using this
expression we obtain
\begin{align}
\sigma^{T}U&(\tau)=\sigma^{T}G_{0}B\tilde{\eta}_{0}\notag\\
&+\sum_{n=1}^{\infty}\biggl[e^{-i\omega_{n}\epsilon}\tilde{\eta}^{T}(BG^{T}\sigma)+e^{i\omega_{n}\epsilon}(\sigma^{T}G^{*}B)\tilde{\eta}^{*}\biggr]~.
\end{align}
The average quantity then can be rewritten as 
\begin{equation}
\langle e^{-\lambda W_{\tau}-i\sigma^{T}U(\tau)} \rangle=\prod_{n=0}^{\infty}\langle e^{s_{n}} \rangle~,
\end{equation}
where
\begin{align}
s_{n}=&-\lambda \tau \tilde{\eta}^{T}c_{n}\tilde{\eta}^{*}+ \tilde{\eta}^{T}\alpha_{n} +\alpha_{-n}^{T}\tilde{\eta}^{*}\notag\\
&-\f{\lambda}{\tau}\f{\gamma}{D}\Delta U^{T}(G^{T}A_{1}G^{*})\Delta U\quad
\text{for}~n\ge 1~,
\end{align}
and 
\begin{equation}
s_{0}=-\f{\lambda \tau}{2}\tilde{\eta}_{0}^{T}c_{0}\tilde{\eta}_{0}+\alpha_{0}^{T}\tilde{\eta}_{0}-\f{\lambda}{2 \tau}\f{\gamma}{D}\Delta U^{T}(G^{T}_{0}A_{1}G_{0})\Delta U,
\end{equation}
in which we have used the following definitions
\begin{align}
c_{n}&=\f{\gamma}{D}BG^{T}A_{1}G^{*}B~,\\
\alpha_{n}&=\lambda \f{\gamma}{D} (BG^{T}A_{1}G^{*})\Delta U-i e^{-i\omega_{n}\epsilon}BG^{T}\sigma~.
\end{align}
We can now calculate the average $\langle e^{s_{n}} \rangle$
independently for each $n\geq1$ with respect to the Gaussian PDF
$P(\tilde{\eta})=\pi^{-2}(\det \Lambda)^{-1}\exp(-\tilde{\eta}^{T}\Lambda^{-1}\tilde{\eta}^{*})$
with $\Lambda^{-1}=\f{2D}{\tau}$diag$(1,\theta/\tau^{2}_{0})$, which
gives, 
\begin{equation}
\langle e^{s_{n}}\rangle=\f{\exp[\alpha_{-n}^{T} \Omega_{n}^{-1}\alpha_{n}-\f{\lambda}{\tau}\f{\gamma}{D}\Delta U^{T}(G^{T}A_{1}G^{*})\Delta U]}{\det(\Lambda \Omega_{n})}~,
\end{equation}
where $\Omega_{n}=\lambda \tau c_{n}+\Lambda^{-1}$. Similarly, calculating the average of $n=0$ term with respect to the Gaussian PDF $P(\tilde{\eta}_{0})=(2\pi)^{-1}(\det \Lambda)^{-1/2}\exp(-\f{1}{2}\tilde{\eta}_{0}^{T}\Lambda^{-1}\tilde{\eta}_{0})$, we get 
\begin{equation}
\langle e^{s_{0}}\rangle=\f{\exp[\f{1}{2}\alpha_{0}^{T} \Omega_{0}^{-1}\alpha_{0}-\f{\lambda}{2 \tau}\f{\gamma}{D}\Delta U^{T}(G_{0}^{T}A_{1}G_{0}^{*})\Delta U]}{\sqrt{\det(\Lambda \Omega_{0})}}~.
\end{equation}
The restricted moment generating function can now be rewritten as 
\begin{equation}
Z(\lambda,U,\tau|U_{0})=\int \frac{d^{2}\sigma}{(2\pi)^{2}} e^{i\sigma^{T}U}\prod_{n=0}^{\infty}\langle e^{s_{n}} \rangle~,
\label{R-GF-2}
\end{equation}
where using the fact $\langle e^{s_{n}} \rangle=\langle
e^{s_{-n}} \rangle$, we can write
\begin{align}
&\prod_{n=0}^{\infty}\langle e^{s_{n}} \rangle=\exp \left
(-\frac{1}{2}\sum_{n=-\infty}^{\infty} \ln[\det(\Lambda\Omega_{n})]\right
) \notag \\
&\times \exp \left(~\frac{1}{2\tau}\sum_{n=-\infty}^{\infty}[\alpha_{-n}^{T}\tau\Omega_{n}^{-1}\alpha_{n}-\lambda\f{\gamma}{D}\Delta U^{T}G^{T}A_{1}G^{*}\Delta U]\right )~.
\label{sum-1}
\end{align}
The determinant in \eref{sum-1} is found to be
\begin{equation}
\det(\Lambda\Omega_{n})=1+\f{4 \theta \lambda(1-\lambda)}{\tau_{0}^{2}\tau_{\gamma}^{2}}|G_{11}|^{2}|G_{22}|^{2}~.
\label{determinant}
\end{equation}
Now in large-$\tau$ limit, we can replace the summations over $n$ into
an integral over $\omega$
i.e. $\sum_{n}\rightarrow\tau\int\f{d\omega}{2\pi}$. The first part of
the summation is then
\begin{equation}
\tau \mu(\lambda)=-\f{\tau}{2} \int \f{d\omega}{2\pi}\ln \Bigl[\det\bigl(\Lambda\Omega(\omega)\bigr) \Bigr]~,
\end{equation}
where $\mu(\lambda)$ is given by \eref{mu}. Similarly, the second part
of the summation can be converted into an integral. Finally, after
doing some manipulations, we obtain
\begin{equation}
\prod_{n=0}^{\infty}\langle e^{s_{n}} \rangle \approx e^{\tau \mu(\lambda)} \exp \biggl[-\f{1}{2}\sigma^{T}H_{1}\sigma 
+i\Delta U^{T}H_{2}\sigma+\f{1}{2}\Delta U^{T}H_{3}\Delta U \biggr]~,
\label{sum-2}
\end{equation}
in which we have defined the following matrices
\begin{align}
H_{1}&=\int_{-\infty}^{\infty} \f{d\omega}{2\pi}G^{*}B(\tau\Omega^{-1})BG^{T}~, \\
H_{2}&=-\lim_{\epsilon\rightarrow 0}\f{\lambda}{2\pi} \f{\gamma}{D} \int_{-\infty}^{\infty} d\omega e^{iw\epsilon}G^{+}A_{1}GB(\tau\Omega^{-1})^{*}BG^{+},
\intertext{and}
H_{3}&=-\f{\lambda}{2\pi}\f{\gamma}{D}\int_{-\infty}^{\infty}
d\omega~G^{T}A_{1}G^{*}\notag \\
&+\f{\lambda^{2}}{2\pi}\f{\gamma^{2}}{D^{2}}\int_{-\infty}^{\infty} d\omega~G^{T}A_{1}G^{*}B(\tau\Omega^{-1})BG^{T}A_{1}G^{*}~.
\end{align}
We then evaluate the matrices by performing the integral by the method of contours. For convenience, we write down the elements of the matrices respectively.
\begin{subequations}
\label{H1-matrix-elements}
\begin{align}
H_{1}^{11}&=\f{D\tau_{\gamma}}{m^{2}}\f{1}{1+\delta \bar{\nu}}\biggl(\delta+\f{1+\theta}{\nu}\biggr), \\
H_{1}^{12}=H_{1}^{21}&=\f{D\theta}{m}\f{1-2\lambda}{\nu(1+\delta \bar{\nu})},  \\
H_{1}^{22}&=\f{D\theta}{\tau_{0}}\f{1}{1+\delta \bar{\nu}}\biggl(1+\f{\delta}{\nu}\biggr)~.
\end{align}
\end{subequations}
The elements of $H_{2}$ matrix are
\begin{subequations}
\label{H2-matrix-elements}
\begin{align}
H_{2}^{11}&=\f{1}{\nu(1+\delta \bar{\nu})}\left[\lambda \theta+\f{1}{2}(1-\nu)+\f{1}{2}\delta\nu(1-\bar{\nu})\right],      \\
H_{2}^{12}&=-\f{\lambda \gamma \theta}{\nu(1+\delta \bar{\nu})},  \\
H_{2}^{21}&=-\f{\lambda \delta}{\gamma \nu(1+\delta \bar{\nu})}+\f{\delta(1-\nu)}{2 \gamma \nu(1+\delta \bar{\nu})},  \\
H_{2}^{22}&=\f{\delta (1-\nu \bar{\nu})}{2 \nu(1+\delta\bar{\nu})}~.
\end{align}
\end{subequations}
\\
The elements of $H_{3}$ matrix are given by
\begin{subequations}
\label{H3-matrix-elements}
\begin{align}
&H_{3}^{11}=\f{\lambda^{2}\theta\gamma^{2}\tau_{\gamma}}{D\nu(1+\delta \bar{\nu})},  \\
&H_{3}^{12}=H_{3}^{21}=\nn\\
&\f{4\lambda^{2}(1-\lambda)\gamma\theta}{D\tau_{0}\tau_{\gamma}} 
\f{1+\nu+(1+\delta\bar{\nu})(1-\bar{\nu}-\f{2}{\delta})}{[1+(1+\delta\bar{\nu})+\delta \nu]\times[1-\f{1}{\delta}(1+\delta\bar{\nu})+\f{\nu}{\delta}]},  \\
&H_{3}^{22}=-\f{\lambda(1-\lambda)\delta\tau_{0}}{D\nu(1+\delta \bar{\nu})} ~.
\end{align}
\end{subequations}

We note that the matrices $H_{1}$ and $H_{3}$ are symmetric and they
satisfy the relation
$H_{3}=(I+H_{2})H_{1}^{-1}H_{2}^{T}$. Inserting \eref{sum-2}
into \eref{R-GF-2} and performing the Gaussian integral over $\sigma$,
we obtain
\begin{align}
Z(\lambda,U,\tau|U_{0})\approx& \f{e^{\tau \mu(\lambda)}}{2\pi\sqrt{\det(H_{1}(\lambda))}} \nn \\
&\times~~e^{-\f{1}{2}U^{T}L_{1}(\lambda)U}~~e^{-\f{1}{2}U_{0}^{T}L_{2}(\lambda)U_{0}}~,
\end{align}
where $L_{1}(\lambda)=H_{1}^{-1}(I+H_{2}^{T})$ and
$L_{2}(\lambda)=-H_{1}^{-1}H_{2}^{T}$. We immediately identify the
right and left eigenfunctions respectively as
\begin{subequations}
\label{psi-chi}
\begin{align}
&\Psi(U,\lambda)=\f{1}{2\pi\sqrt{\det(H_{1}(\lambda))}}\exp \biggl[-\f{1}{2}U^{T}L_{1}(\lambda)U \biggr]~,  \\
&\chi(U_{0},\lambda)=\exp \biggl[-\f{1}{2}U_{0}^{T}L_{2}(\lambda)U_{0} \biggr]~.
\end{align}
\end{subequations}
It is then straightforward to verify
 $\mathcal{L}_{\lambda}\Psi(U,\lambda)=\mu(\lambda)\Psi(U,\lambda)$
 and $\int dU \chi(U,\lambda) \Psi(U,\lambda)=1$.  The steady state
 distribution is given by
\begin{align}
P_\mathrm{SS}(U)&=Z(\lambda=0,U,\tau\rightarrow \infty|U_{0})=\Psi(U,\lambda=0) \nn \\
&=\f{1}{2\pi\sqrt{\det(H_{1}(0))}}\exp \biggl[-\f{1}{2}U^{T}L_{1}(0)U \biggr]~,
\label{steady-state}
\end{align}
where $L_{1}(0)$ and given by
\bea
L_{1}(0) &= \f{1}{\det{H_{1}(0)}}\f{D}{1+\delta} 
\begin{pmatrix}
\f{\theta}{\tau_{0}}(1+\delta) & -\f{\theta}{m} \\
-\f{\theta}{m} & \f{\tau_{\gamma}}{m^{2}}(1+\delta+\theta)\\
\end{pmatrix}.
\label{L_{1}(0)}
\eea
It is worth noting that 
the deviation of the system from equilibrium can also be measured 
 using \eref{steady-state}
\begin{equation}
\alpha=\f{\langle v^{2}\rangle_\mathrm{ss}}{\langle v^{2}\rangle_\mathrm{eq}}-1~,
\label{alpha}
\end{equation}
where $\langle v^{2}\rangle_\mathrm{ss}$ is the velocity variance in the
steady state which can be found from \eref{L_{1}(0)} and $\langle
v^{2} \rangle_\mathrm{eq}$ is that of in equilibrium in the absence of the
external driving. Hence, one finds, $\alpha=\theta/(1+\delta)$.  

Now, averaging the restricted generating function with respect to the
steady state distribution $P_\mathrm{SS}(U)$, we get
back \eref{Z-asymptotic}, where $g(\lambda)$ is given by
\begin{align}
g(\lambda)=\bigl[\det(I+H_{2}^{T})\bigr]^{-1/2} 
\bigl[\det(I-H_{1}(0)H_{1}^{-1}(\lambda)H_{2}^{T}(\lambda))\bigr]^{-1/2}~,
\label{glambda-2}
\end{align}
where the first and second terms are due to tracing out the final and initial variables respectively. Using the forms of the 
matrices given by \eref{H1-matrix-elements}
and \eref{H2-matrix-elements}, we obtain 
\begin{subequations}
\label{glambda-3}
\begin{align}
f_{1}(\lambda,\theta,\delta):&=\det(I+H_{2}^{T})\nn\\&=\f{1}{4\nu
(1+\delta\bar{\nu})^{2}}\Bigl[p(\lambda) + 2\theta \lambda q(\lambda)\Bigr]
, \\
f_{2}(\lambda,\theta,\delta):&=\det[I-H_{1}(0)H_{1}^{-1}(\lambda)H_{2}^{T}(\lambda)]
\nn\\
&=\f{1}{4(1+\delta)^{2}}\f{1}{\theta+(1+\delta\bar{\nu})^{2}}
\Bigl[r(\lambda) + 2\theta\lambda s(\lambda)\Bigr].
\end{align}
\end{subequations}
where
\begin{subequations}
\begin{align}
p(\lambda)=&2 + 2\nu+\delta~(1+\bar{\nu})~(1+\delta+3\nu+\delta\nu\bar{\nu}), \\
q(\lambda)=&2+\delta(\bar{\nu}-1)
=1+\sqrt{1+\delta^{2}+2\delta\nu}-\delta.
\end{align}
\end{subequations}
and 
\begin{subequations}
\begin{align}
r(\lambda)&=2\theta(1+\nu)+2(1+\nu)(1+\delta)^{2} \nn \\
&+ \Bigr[\theta+(1+\delta)^{2} \Bigl]\Bigr[\delta(1+\bar{\nu})^{2}+\delta(1+\bar{\nu})(1+\delta\bar{\nu})(\nu+\bar{\nu})\Bigl]~,\\
\label{s}
s(\lambda)&=-\bigl[2+2\theta+3\theta\delta + \delta\bar{\nu}
+ \theta\delta\bar{\nu} \bigr]\nn \\
&\quad+ \bigl[\delta+2\delta^{2}(2+\bar\nu)+\delta^{3} (1+3\bar\nu)
\bigr]~.
\end{align}
\end{subequations}

Let us now analyze the functions $f_{1}(\lambda,\theta,\delta)$ and
$f_{2}(\lambda,\theta,\delta)$ in details.  We note that the
pre-factors outside the square bracket of
$f_{1}(\lambda,\theta,\delta)$ and $f_{2}(\lambda,\theta,\delta)$ are
always positive. Moreover, $p(\lambda)$ and $q(\lambda)$ are again
clearly positive in the region
$\lambda\in[\lambda_{-},\lambda_{+}]$. In particular, they take the
minimum values at $\lambda_\pm$, given by $p(\lambda_\pm) = 2 + a_1$
and $q(\lambda_\pm)= 1+ a_2=2-a_3$, where
$a_1=(1+\delta)(\delta+\sqrt{1+\delta^2}-1) \ge 0$, $1 \ge
a_2=\sqrt{1+\delta^2}-\delta > 0$, and
$1>a_3=(1+\delta)-\sqrt{1+\delta^2} \ge 0$. Therefore,
$f_1(\lambda_+,\theta,\delta) > 0$ as $\lambda_+ >0 $.  On the other
hand, at $\lambda=\lambda_-$ we get
\begin{align*}
p(\lambda_-) + 2 \theta\lambda_-\,  q(\lambda_-) 
&=(2+a_1) +  2\theta\lambda_- (2-a_3)\\
&=a_1 + (- 2 a_3 \theta \lambda_-) + 2(1+2\theta\lambda_-). 
\end{align*}
The first two summands in the last line of the above expression is
clearly positive (note that $\lambda_- <0$). Moreover, it can be shown that
\begin{equation}
1+2 \theta \lambda_{-}=
\sqrt{1+\theta}\bigl[\sqrt{1+\theta}-\sqrt{\theta}\bigr] >0.
\label{condition-1}
\end{equation}
This also implies that
\begin{equation}
1+2 \theta \lambda >0 \quad \text{for}~\lambda\in[\lambda_-,\lambda_+].
\label{condition-1b}
\end{equation}
 Therefore, $f_1(\lambda_-,\theta,\delta) >0$, which implies that
$f_{1}(\lambda,\theta,\delta)$ stays positive in the
region~$\lambda\in[\lambda_{-},\lambda_{+}]$.  

Similarly, we can analyze the second term
$f_{2}(\lambda,\theta,\delta)$.  Clearly, $r(\lambda)$ is always
positive in the region $\lambda\in[\lambda_{-},\lambda_{+}]$. On the
other hand, the first line in the expression of $s(\lambda)$ given
by \eref{s} is negative whereas the second line is positive;
$s(\lambda)$ can take both positive and negative values in the
$(\theta, \delta, \lambda)$ space.  
Writing \eref{s} as $s(\lambda)= -b_1 + b_2$ with both $b_1>0$ and
$b_2 > 0$, we get
\begin{equation*}
r(\lambda) +
2 \theta \lambda s(\lambda)= \bigl[r(\lambda) - b_2\bigr] + (1+2\theta\lambda) b_2 +
(-2b_1\theta\lambda).
\end{equation*}
By explicitly expanding $r(\lambda)$, it can be seen that all the
terms appearing in $b_2$ completely cancel with some of the terms of
$r(\lambda)$. Therefore, $r(\lambda) - b_2 > 0$ for $\lambda\in
[\lambda_-,\lambda_+]$. Similarly, according to \eref{condition-1b},
the second summand is positive. Finally, the last summand is clearly
positive for $\lambda <0$. Therefore, 
 $f_{2}(\lambda,\theta,\delta)>0$ for $\lambda_- \le \lambda \le 0$.

At $\lambda=\lambda_{+}$, we find that $r(\lambda_+)+2\theta \lambda_+
s(\lambda_+)$ changes sign in the parameter space of
$(\theta,\delta)$.  The phase boundary that separates the two regions
where this function stays positive and negative respectively is given
by
\begin{equation}
f_{2}(\lambda_{+},\theta,\delta)=0~,
\label{phase-curve}
\end{equation}
which is shown in  \fref{phase-space}.


\end{document}